# Adhesion mechanics of graphene membranes

J.S. Bunch *, M.L. Dunn

*Department of Mechanical Engineering, University of Colorado, Boulder, CO*

*80309 USA*

*Corresponding author: jbunch@colorado.edu (J.S. Bunch)

*Abstract*

The interaction of graphene with neighboring materials and structures plays an important role in its behavior, both scientifically and technologically. The interactions are complicated due to the interplay between surface forces and possibly nonlinear elastic behavior. Here we review recent experimental and theoretical advances in the understanding of graphene adhesion. We organize our discussion into experimental and theoretical efforts directed toward: graphene conformation to a substrate, determination of adhesion energy, and applications where graphene adhesion plays an important role. We conclude with a brief prospectus outlining open issues.

## 1. INTRODUCTION

Graphene has attracted significant interest and excitement due to its unprecedented mechanical, electrical, and thermal properties, as well as the attraction of creating a graphene-based device manufacturing infrastructure [1,2]. At the same time, graphene and other materials, structures, and devices are increasingly influenced by surface forces as their size moves into the nm-range. This occurs because i) the materials are often separated by small distances and are sensitive to the operant range of surface forces, ii) the structural stiffness decreases as its size decreases, and iii) both the surface forces and structural stiffness scale nonlinearly with relevant dimensions. For example, van der Waals energy between two molecules varies with separation, $d$, as $1/d^6$ over the range of ~1-10 nm and then transitions to $1/d^7$ for separations $d > $ ~100 nm due to retardation effects, and the bending stiffness of a beam varies with thickness, $t$, as $t^3$.

Adhesion plays an important role in many important technological applications of graphene. For example, adhesive interactions are critical to nanomechanical devices [3]. For example, graphene switches are actuated electrostatically to bring them into, or near, contact with an electrode but van der Waals forces can permit the release of the switch [4–6]. A promising route to manufacture graphene involves CVD growth on an appropriate substrate followed by batch transferring it from the host substrate to a functional target substrate for device applications [7]. The engineering of the peeling, stamping, and other fabrication processes depends on the adhesion of the graphene to both the host and target substrates.

Scientifically, graphene offers exciting avenues to approach important questions related to surface forces. Because it is the ultimate thin membrane, graphene can conform more closely to a surface than any other solid. This provides new opportunities to study solid-solid surface interactions including the effects of even the smoothest surface topographies and potentially the nature of van der Waals and Casimir forces.

Here we review recent research under the broad theme of graphene adhesion, organizing our discussion into experimental and theoretical efforts directed toward understanding graphene conformation to a substrate, determining adhesion energy, and applications of graphene adhesion. We note that in many regards graphene adhesion can be considered within the broader context of ultrathin membrane adhesion, but our focus here is only on graphene [8,9]. We conclude with a brief prospectus on interesting open issues in the field.

## 2. EXPERIMENTS ON GRAPHENE ADHESION

Approaches to determine the adhesion energy of graphene and a substrate typically involve experiments where graphene in adhesive contact with a substrate is delaminated from it by well-controlled forces or the deformation of graphene conforming to well-defined surface features is measured. Both measurements use an appropriate mechanics model that describes the balance between adhesion energy and strain energy at equilibrium. These types of experiments are reviewed in this section.

### 2.1. Graphene Conformation

The first mechanical devices made from graphene were mechanical resonators fabricated by exfoliation of graphene over predefined trenches to form doubly clamped beam resonators [3] with mechanical resonant frequencies of in the ~10 - 150 MHz range (Fig. 1a). This suggested that the graphene doubly clamped beams were subject to tension on the order of nNs, the origin of which was unclear until the first membrane/drum resonators were fabricated and measured by Bunch et al. [10]. This work showed self-tensioning in the graphene resonators due to adhesion to the sidewalls by the same van der Waals forces that clamp the graphene resonators to a substrate. The magnitude of the tension (~0.1 N/m) was deduced from the resonant frequency and verified by AFM indentation experiments [10]. Further experiments by Barton et al on graphene drum resonators fabricated by chemical vapor deposition (CVD) showed this tension to be fairly consistent over a large range of graphene membrane diameters [11]. The adhesion to the sidewalls was also seen by Lee et. al, though the magnitude was an order of magnitude smaller than that measured by Bunch et al. [12]. Conformation of initially flat graphene along the sidewalls resulting from adhesion energies of ~ 0.1 J/m$^2$ should be < 1nm [13,14]. Dips of several nanometers as measured by AFM by both groups may suggest considerable slack; however, such slack should show up as large upward deflections in over pressurized μm size graphene membranes. This is inconsistent with experiments by Koenig et al which show little slack in over pressurized exfoliated graphene membranes [15]. Even though experiments on a wide range of suspended graphene devices suggest a strong self-tensioning, the large dips seen by AFM in exfoliated graphene membranes are inconsistent with the small strains and small slack in these devices suggesting instead that the large dips might be an imaging artifact due to the AFM tip imaging a flexible suspended membrane.

The ability of graphene to conform extremely well to a substrate is clear from scanning probe microscopy studies of graphene on varying substrates [16–19]. The roughness of a graphene-on-substrate configuration will depend on how well the graphene conforms to the rough surface and it can in principle be tuned and made smoother than the substrate. Measurements on SiO$_x$ and mica substrates [16] show that graphene on SiO$_x$ had an rms roughness of $\sigma = 154$ pm with a correlation length of $l = 22$ nm (compared to the underlying SiO$_2$ substrate with $\sigma = 168$ pm, $l = 16$ nm) and graphene on mica had $\sigma = 24.1$ pm, $l = 2$ nm (compared to the underlying mica substrate with $\sigma = 34.3$ pm, $l = 2$ nm) (Fig. 1c and 1d). The interplay between the graphene-substrate adhesion energy and the graphene deformation plays a key role in determining the equilibrium graphene conformation.

Scharfenberg et al. (2011) mechanically exfoliated graphene onto a PDMS substrate with one-dimensional sinusoidal corrugations on its surface of 1.5 μm wavelength and 200 nm depth [20]. Because it is so soft, the PDMS corrugated surface also deforms and the elastic energy stored in it contributes to the global energy balance that yields the adhesion energy. Scharfenberg et al. found that the graphene was highly conformal to the corrugated PDMS substrate and by measuring the graphene thickness and deformed graphene with an AFM and combining it with the mechanical analysis they deduced an adhesion energy of 0.07 J/m$^2$. They also showed that for a small number of graphene layers the films would intimately conform to the periodic surface, while for a large number of layers, the graphene would essentially rest on top of the top of the sinusoidal crests [21]. For the experimental parameters mentioned above, a transition between the two states occurred at 61 layers.

## 2.2. Adhesion Energies

Zong et al. (2010) mechanically exfoliated graphene on top of a SiO$_x$ surface covered with ~50-80 nm diameter gold and silver nanoparticles (Fig. 2a) [22]. The graphene adhered to the SiO$_x$ but was draped over the nanoparticles leaving a circular blister between the graphene and SiO$_x$ in a region surrounding the nanoparticle. They measured the particle height $w$, blister radius $a$, and graphene thickness $h$ with an AFM. A membrane mechanics model then provided the adhesion energy through $\gamma = (Ehw^4)/(16a^4)$. Zong et al. (2010) took $E = 0.5$ TPa, resulting in a graphene-SiO$_x$ adhesion energy of 0.15 J/m$^2$; if they would have used E ~1 TPa, consistent with more recent theory and measurements, the adhesion energy would be 0.3 J/m$^2$ [10,12]. Two challenges with this approach are i) the potential expansion of gas trapped in the circular blister during testing in a high-vacuum SEM chamber which would increase $a$, meaning that the actual adhesion energy could be higher, and ii) measurement of the graphene thickness with AFM for single or very few layer graphene.

Koenig et al. (2011) determined the graphene-$SiO_x$ adhesion energy for 1-5 layers of graphene using a pressurized blister test with graphene sheets on a $SiO_2$ substrate patterned with circular microcavities (5 µm diameter, 300 nm depth) [15]. They placed chips with multiple microcavities in a high pressure nitrogen chamber until the $N_2$ gas was equilibrated inside and outside of the microcavity at a prescribed pressure and then removed them to ambient conditions (Fig. 2b). This results in a pressure difference across the graphene membrane that causes it to bulge and as a result increases the microcavity volume and decreases the $N_2$ pressure (Fig. 2c). At a large enough pressure the membrane will delaminate from the substrate in a stable manner because the number of $N_2$ molecules is constant during the process. Koenig et al systematically increased the charging pressure and measured bulged and delaminated graphene membrane shapes with an AFM (Fig. 2d). They directly measured the elastic properties of the graphene with the bulge test and determined its thickness by Raman measurements. They coupled the measurements with a mechanics analysis of the blister configuration to determine a graphene-$SiO_x$ adhesion energy of 0.45 J/m$^2$ for a single layer of graphene and 0.31 J/m$^2$ for multilayers of 2-5 layers. The reason for the difference between 1 and 2-5 graphene layers was speculated to be a result of varying levels of conformation of the graphene membrane with the $SiO_x$ substrate roughness as a function of the number of graphene layers.

Recently Yoon et al directly measured the adhesion energy of CVD grown graphene on copper using a traditional peel test (Fig. 2f) [23]. In these experiments, graphene is glued to a target substrate with epoxy and a force is applied to delaminate the graphene from the copper substrate. Adhesion energies of 0.72 J/m$^2$ were found. This is considerably higher than previous measurements on adhesion energy between graphene and a substrate, and the authors attributed this large adhesion energy to the possible increase in the electronic density at the interface between graphene and the copper surface.

## 2.3. Experimental Applications of Graphene Adhesion

Measured graphene adhesion is strong compared to typical micromechanical structures [24,25] and in many cases this is detrimental to device performance. One example is graphene nanomechanical switches where suspended graphene is brought into contact with an underlying electrode [3–6] (see, e.g., Fig. 3a). These early nanomechanical switches suffer from strong adhesive forces that are detrimental to device performance and future work is needed to engineer the graphene substrate interface to minimize adhesive forces. Similar stiction problems haunted the development of MEMS devices such as the micromirrors found in Texas Instruments DLP projectors and it took years of research to overcome them before this technology finally made it to the market. Graphene switches face a similar daunting task but research is still at its infancy and the promise of a one atom thick conducting material with such a low bending rigidity acting as the switching element is promising.

Another example where adhesion engineering is important is graphene manufacturing which requires graphene films to be grown and transferred to suitable substrates. Yoon et al demonstrated the transfer of graphene by reproducibly peeling it from a copper substrate and transferring it to another suitable surface. Li et al. did a theoretical and experimental study of graphene stamping onto an $SiO_x$ substrate from a graphite host crystal [26]. In both of these cases, a thorough knowledge of the adhesion energies between graphene and varying substrates is critical.

Sliding and friction is another case where adhesion plays an important role. Recent measurements of the friction of atomically thin materials with an AFM tip was showed a strong layer dependence [27]. The importance of sliding was seen in experiments by Conley et al who used bimetallic like cantilevers from CVD graphene films transferred to microfabricated cantilevers and estimated the interfacial shear strength to be ~ 1GPa[28] and by Liu et al who measured the shear modulus of monolayer graphene and found it to be 5 x larger than multilayer graphene [29]. Interestingly, Conley et al found strong temperature dependence in the interfacial shear strength which approaches 0 for temperatures ~ 500K. The shear strength and adhesion energy are closely related since they both depend on the short range forces between the neighboring atoms. This temperature dependence in the interfacial shear strength suggests yet another route to engineer the adhesion energy of graphene.

A recent development in graphene physics where high adhesion energies would be beneficial is in strain engineering the electronic properties of graphene or straintronics [30,31]. In this case, large strains are used to deform graphene and modify the electronic band structure. Evidence of graphene nanobubbles with such large trains localized on a nm scale are seen in wrinkled graphene films grown on Pt[31]. The strains typically needed for strain effects to influence the graphene bandstructure significantly are > 5%, much larger than what is currently accessible for pressurized graphene blisters as seen in Fig. 2. The maximum strain available for over pressurized graphene balloons is limited by the adhesion energy, and to reach a maximum strain of ~5%, regardless of the bubble diameter, requires an adhesion energy of ~ 3 J/m$^2$ [32]. Realizing such large strains in pressurized suspended graphene membranes without delamination is challenging due to the large adhesion energies needed. However, other creative geometries that can induce such large strains over larger areas of graphene may prove viable paths to enable graphene straintronics (Fig. 4).

### III. THEORY AND MODELING GRAPHENE ADHESION

Theory and modeling efforts to understand adhesion of graphene have focused on two general areas: i) the mechanics of a thin membrane adhered to a substrate based on experimentally realizable geometrical configurations, and ii) the influence of surface forces, graphene thickness (number of layers), and substrate roughness [14,33–40] on the effective adhesion energy of the graphene/substrate pair. The former are typically used to model experimental configurations and then used inversely for the extraction of the effective adhesion energy. Here we focus on the latter.

Sasaki et al. (2009) simulated peeling of a graphene monolayer from a flat, rigid graphite substrate using molecular mechanics and described the graphene-substrate van der Waals interactions by a Lennard-Jones potential [41]. They started with a rectangular graphene membrane adhered to the substrate, pulled a region of atoms at the center of the graphene away from the substrate, and recorded the force-displacement response as well as the resulting configurations. They found that graphene peels from the surface in successive partial steps around the load point that appear as discrete jumps in the force-displacement response. Lu and Dunn (2010) modeled a similar peeling configuration with molecular mechanics and considered configurations of peeling from the sidewalls of a cavity, like those observed experimentally. In addition, they developed theory to describe the pretension that can occur due to adhesion, peeling, and sliding of the graphene [10,14]. They obtained excellent agreement between theory and atomistic simulations and identified the influence of van der Waals adhesion energy, membrane elasticity, geometry, and loading on graphene peeling from and/or sliding along a substrate.

While these studies focused on adhesion to a flat substrate, a recent series of papers have addressed the effects of a surface roughness, which always exists in reality. To help understand the basic phenomena, consider a graphene sheet in adhesive contact with a rough substrate. The graphene will assume an equilibrium configuration where it conforms to the rough surface to a degree dictated by the balance between the energy of the surface forces and the elastic energy stored in the graphene due to local deformation (bending and possibly stretching and sliding). The *effective adhesion energy*, as would be measured in any experiment that peels the graphene from the substrate, is influenced by the topographical conformation between the graphene and the rough surface; this can result in an actual *adhesion area* that exceeds the nominal surface area. The relevance of various factors on the adhesion of a thin plate (graphene) with Young's modulus $E$ to a rough surface can be described to leading order by an adhesion parameter [42]:

$$\eta = \frac{E t^3 h^2}{\Gamma \lambda^4} \qquad (1)$$

In Eq. (1) Γ is the adhesion energy between graphene and a flat substrate and η describes the competition between Γ and the elastic energy of the deforming plate based on a simple analysis of the energy contributions required for the plate to conform to an idealized rough surface described by asperities of height $h$ and spacing $\lambda$. For $h \ll 1$ the plate can conform to the rough surface while for $h>1$ it can only partly conform.

Recent continuum theory treats a graphene sheet adhered to a rigid or deformable half-space with a surface profile that represents actual or idealized roughness [20,21,33–35,43]. Details of how graphene interacts with a rough substrate require it to be modeled as a plate (with bending and possibly stretching), rather than a membrane. An adhered graphene configuration can be described by a spatially-varying displacement field $w(x,y)$ that has associated with it an elastic strain energy function $U_e(w,D,E,\nu,t)$ where $D$ is the bending modulus, $E$, $\nu$ are Young's modulus and Poisson's ratio and $t$ is the thickness. The surface forces are represented via an interaction energy $U_{int}(V(r), \rho_g, \rho_s)$ where $\rho_g$ and $\rho_s$ are atomic densities of atoms in the graphene sheet and the substrate and $V(r)$ is the interaction potential between atoms of the substrate and graphene, e.g., of the Lennard-Jones 6-12 type. For a prescribed substrate surface profile, the equilibrium configuration of the graphene sheet and the effective adhesion energy are determined by minimizing the sum of the elastic strain energy and interaction energy. Simulations of this kind are challenging if the roughness profile is complicated due to the nonlinear, multibody interaction potential, and the many local minima that can exist in the overall energy landscape.

A series of recent studies consider an idealized scenario with roughness described by a one-dimensional sinusoid (with amplitude $A$ and wavelength $\lambda$) and neglect stretching of the graphene sheet. In this case analytical, although complex, solutions can be obtained for equilibrium configurations and the effective adhesion energy [33,35,43]. The main results of these studies are that the equilibrium membrane configuration, which describes how it conforms to the rough substrate, and the effective adhesion energy, depend on the interplay among the membrane thickness and stiffness and the wavelength and amplitude of the substrate roughness. For a given $A$ and $\lambda$, the extreme case of a thick, stiff membrane will essentially rest on top of the substrate without deforming significantly, while a thin, compliant membrane will significantly bend to conform closely to the substrate. The effective adhesion energy depends on the degree of conformation of the membrane to the substrate, e.g., as characterized by the ratio of membrane to substrate roughness amplitudes. The degree of conformation varies between the limits for a thick and thin membrane, but interestingly not smoothly. A jump can exist in the degree of conformation, and thus the adhesion energy, with system parameters due to instabilities that arise from the interplay from the nonlinear interaction force and linear bending behavior, similar to jump to contact phenomena observed in many surface phenomena.

Koenig et al. (2011) suggested this phenomenon as a possible explanation for the discrepancy between their monolayer and multiple layer graphene adhesion measurements [15]. Their AFM measurements of graphene roughness on a $SiO_x$ substrate showed a decreasing roughness with increasing layer number (~197 pm for bare $SiO_x$, 185 pm with one layer, and 127 pm with 15 layers of graphene) suggesting that monolayer graphene conforms more closely to the $SiO_x$ substrate. They modified the theory described above to account for effects of multilayer graphene, and found that it supports the suggestion of a jump to contact that results in increased adhesion energy as the number of layers decreases, however, they caution that the model of a sinusoidal roughness is too simple to quantitatively predict that the details of their experiments. Another question brought up by such experiments is the origin of a higher adhesion energy between 1 layer and 2-5 layers. Raman measurements on pressurized graphene also suggested a similar increase in the degree of conformation but in their case, between 2 and 3 layers [44]. Gao and Huang recently developed a continuum model that shows the adhesion jump with varying thickness by modeling the bending rigidity of graphene to account for its non-standard dependence on thickness (as $t^3$) as described by Koskenin and Kit (2010) [43,45]. A similar deviation from continuum mechanics for the bending rigidity of graphene was seen in a recent computational study by Zhang et al. [46]. This model, while it demonstrates the qualitative phenomena, is also probably too idealistic to quantitatively describe the results of Koenig et al. In addition to the degree of conformation, another possible explanation for the increased adhesion energy in monolayer graphene is the possibility of chemical bonding of graphene to an $SiO_x$ substrate. Overall, further work is needed to clarify the exact origin of the observed increase in adhesion with decreasing graphene thickness.

## IV. OUTLOOK AND CONCLUSIONS

Our discussion of recent studies of the adhesion of graphene to a substrate highlights many of the important issues that arise with graphene, and more broadly with nanoelectromechanical systems (NEMS). A number of additional, important issues are being pursued both from experimental and theoretical perspectives, and we briefly describe some of them here. Additional unique issues arise in graphene adhesion due to the nature of multilayer graphene where each layer is weakly bonded by van der Waals forces that can be on the same order of magnitude as those with a substrate. In experiments with multilayer graphene it is difficult to determine if there is sliding at the graphene/substrate interface or between graphene layers during delamination [15]. Understanding the possible sliding that may occur is of interest and important; it is related to recent frictional studies between graphene layers and graphene substrate interfaces [27]. The adhesion between individual graphene layers is also important for device manufacturing based on mechanical exfoliation and transfer of graphene [47]. Understanding the effects of surface roughness on adhesion of single and multilayer graphene is in its infancy; while much of the general theory is in place, only simulations of highly-idealized roughness profiles have been performed. These provide some qualitative insight into possible mechanisms, including instabilities and pinning at asperities [37], but are inadequate to describe actual experiments.

Nanomechanical structures, especially graphene, provide an attractive vehicle to study not only adhesion energy but details of the operant surface forces because of their high sensitivity to the weak forces and the inherent difficulties in isolating them from other stronger forces such as electrostatic. Exploration of the long range van der Waals or Casimir forces that arise with graphene are ripe for future experimental and theoretical work. There is recent theoretical work showing that the long range forces are greatly reduced for graphene membranes [48,49] and experimental validation of these results would be welcome. The long range forces are going to play an important role in the development of mechanical devices such as graphene nanomechanical switches.

In conclusion, the interaction of graphene with a substrate or other materials, structures, or devices is being studied intensively around the world from theoretical and experimental perspectives, but the tip of the surface has only been scratched in this exciting field.

**ACKNOWLEDGEMENTS**

The authors thank Steven Koenig, Narasimha Boddetti, Xinghui Liu, Jianliang Xiao and Paul McEuen for useful discussions. The authors are grateful for support by NSF Grants #0900832(CMMI: Graphene Nanomechanics: The Role of van der Waals Forces), #1054406(CMMI: CAREER: Atomic Scale Defect Engineering in Graphene Membranes), the DARPA Center on Nanoscale Science and Technology for Integrated Micro/Nano-Electromechanical Transducers (iMINT), and the National Science Foundation (NSF) Industry/University Cooperative Research Center for Membrane Science, Engineering and Technology (MAST).

**FIGURES**

**FIGURE CAPTIONS**

Figure 1

a) Colorized scanning electron microscope image of a suspended graphene resonator (scale bar = 1 µm) (adapted with permission from Ref. [3] Copyright (2007) American Association for the Advancement of Science). b) Atomic force microscope image of a suspended graphene drum resonator with dimensions 4.5 µm x 4.5 µm. (adapted with permission from Ref. [10] Copyright (2008) American Chemical Society) c-d) Atomic force microscope image of graphene on an c) $SiO_x$ substrate and d) on mica substrate. The image size is 100 nm x 100 nm and maximum z scale is 0.4 nm (adapted with permission from Ref. [16] Copyright (2009) Nature Publishing Group). e) Schematic of a graphene flake conforming to a corrugated substrate (adapted with permission from Ref. [21] Copyright (2012) American Institute of Physics). f) Schematic of a graphene flake unable to conform to a corrugated substrate (adapted with permission from Ref. [21] Copyright (2012) American Institute of Physics). g) Atomic force microscope image of a many layer graphene flake conforming to a corrugated substrate (adapted with permission from Ref. [21] Copyright (2012) American Institute of Physics). Scale bar is 1.5 µm.

Figure 2

a) Schematic showing a graphene flake conforming over a blister perturbation on a silicon substrate (adapted with permission from Ref. [22] Copyright (2010) American Institute of Physics). b) Schematic of an over pressurized suspended graphene membrane (adapted with permission from Ref. [50] Copyright (2011) Nature Publishing Group). c) Atomic force microscope image of a pressurized graphene membrane (adapted with permission from Ref. [15] Copyright (2011) Nature Publishing Group). d) Atomic force microscope line cuts through a center of the pressurized graphene membrane in (c) at varying pressure difference (adapted with permission from Ref. [15] Copyright (2011) Nature Publishing Group). e) Adhesion energy for graphene membranes on a $SiO_x$ substrate (adapted with permission from Ref. [15] Copyright (2011) Nature Publishing Group). f) Adhesion energy measurements on CVD grown graphene on copper measured using a peel test (adapted with permission from Ref. [23] Copyright (2012) American Chemical Society).

Figure 3

a) Schematic of an all graphene electromechanical switch (adapted with permission from Ref. [4] Copyright (2009) American Institute of Physics). b) Schematic showing a configuration of graphene on a substrate with trenches and wells for graphene straintronics (adapted with permission from Ref. [30] Copyright (2009) American Physical Society).

Figure 4

Adhesion energy vs. number of layers for adhesion of multilayer graphene to a surface with sinusoidal topography of a fixed wavelength and various amplitude, d. The results show a jump to adhesion resulting in a jump in the adhesion energy as the number of layers change (adapted with permission from Ref. [43] Copyright (2011) IOP Publishing).

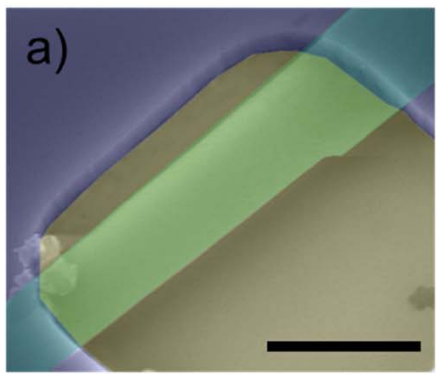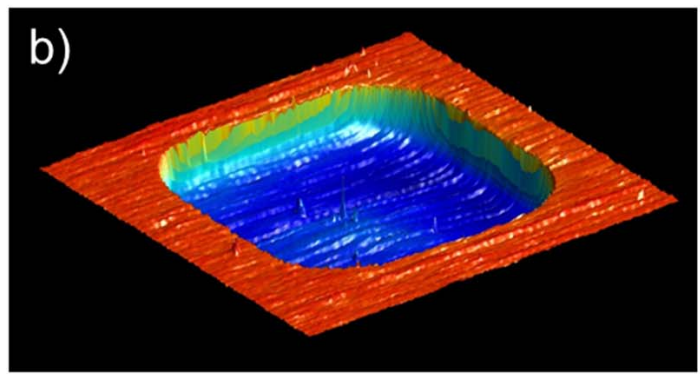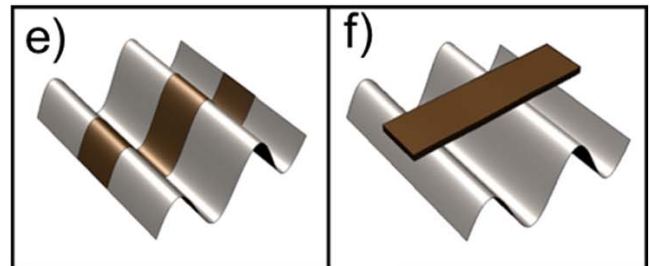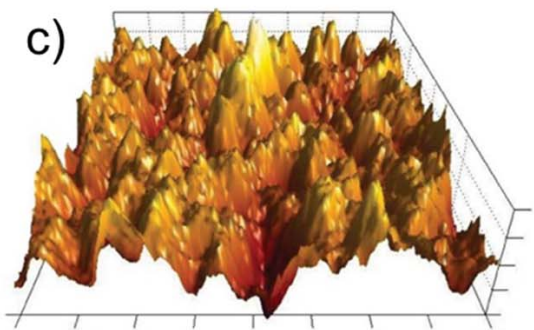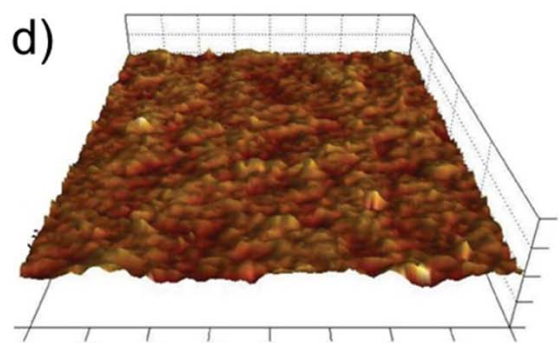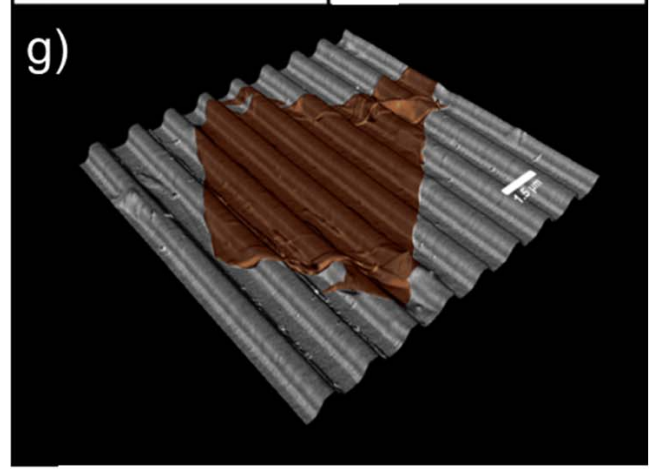

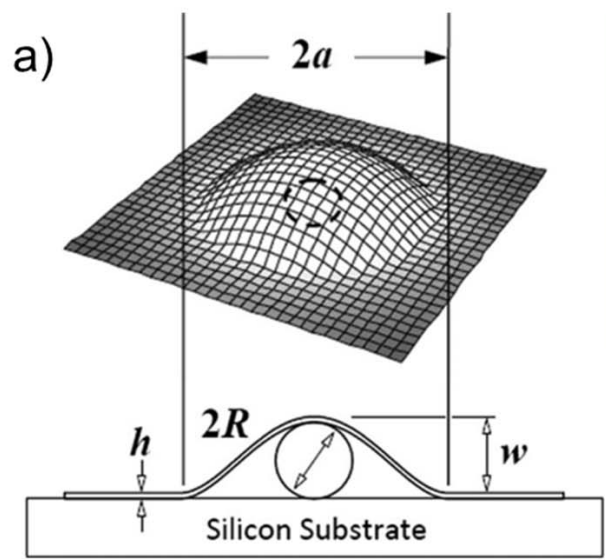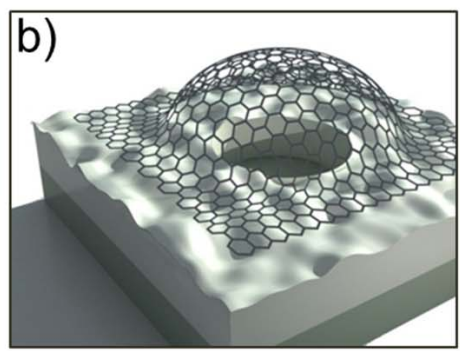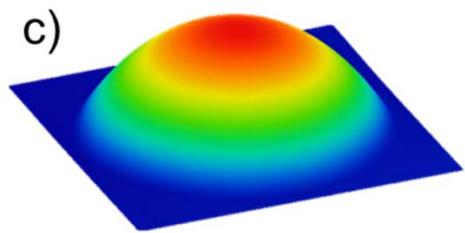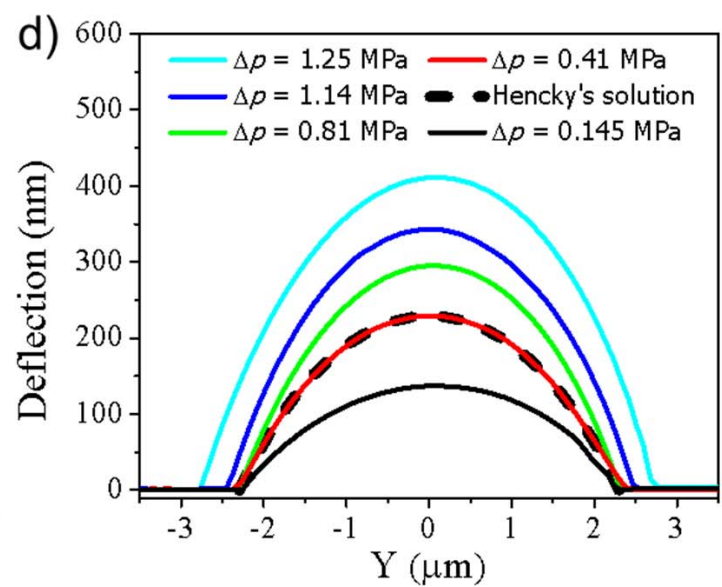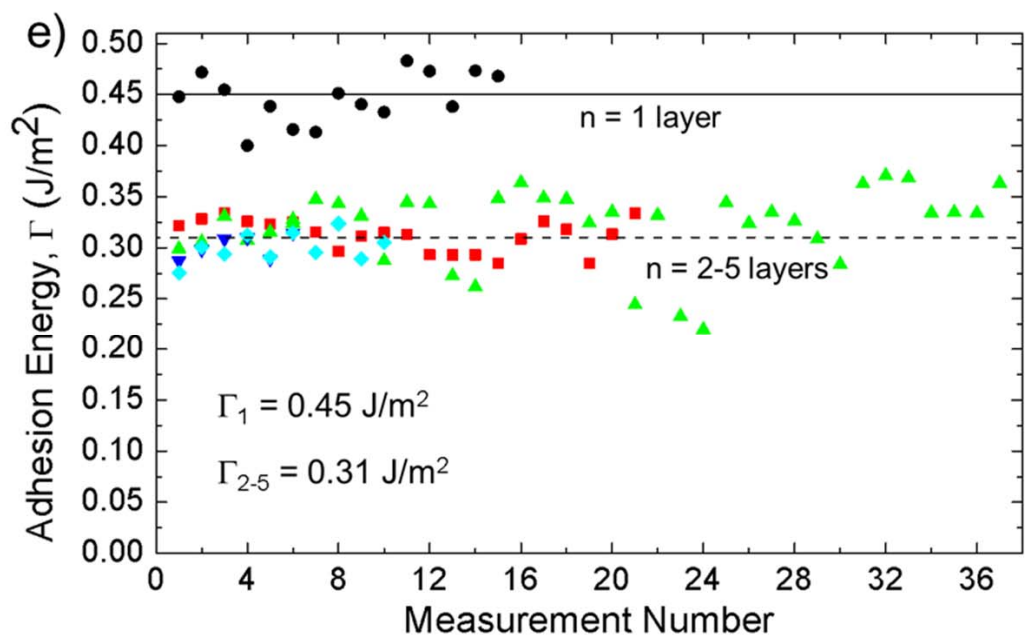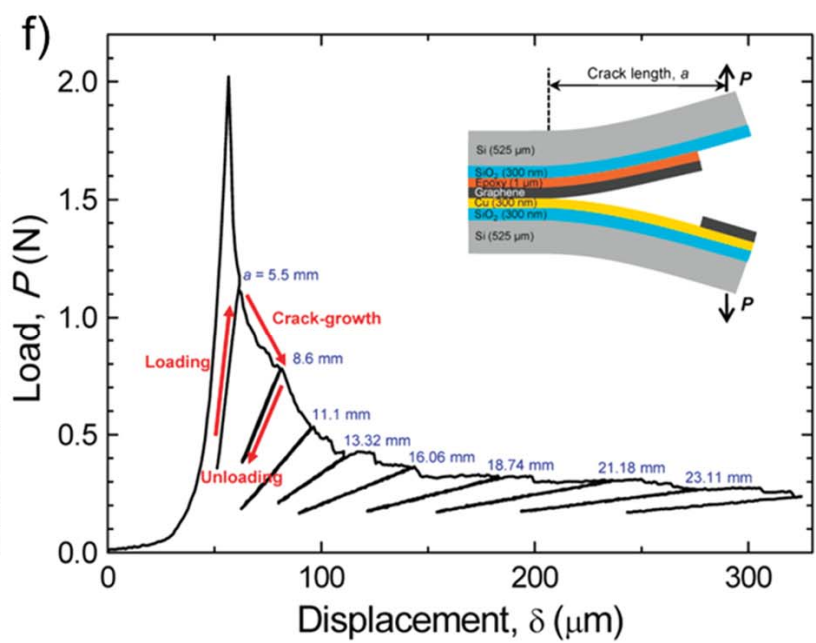

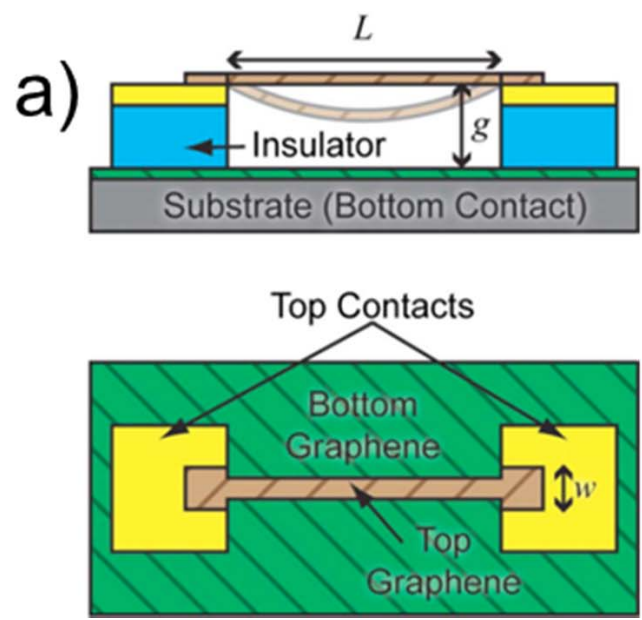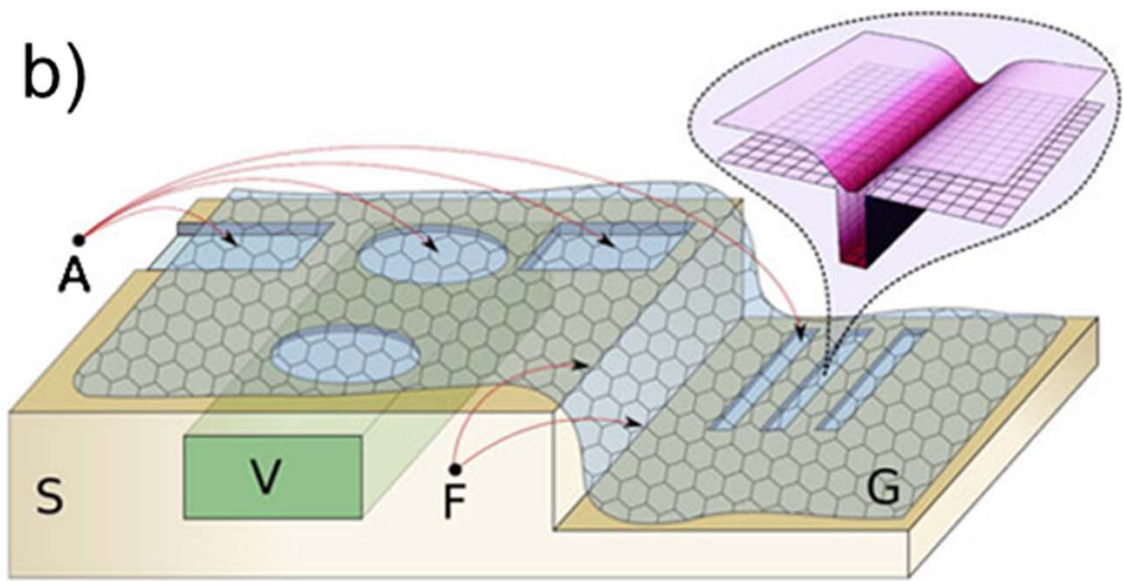

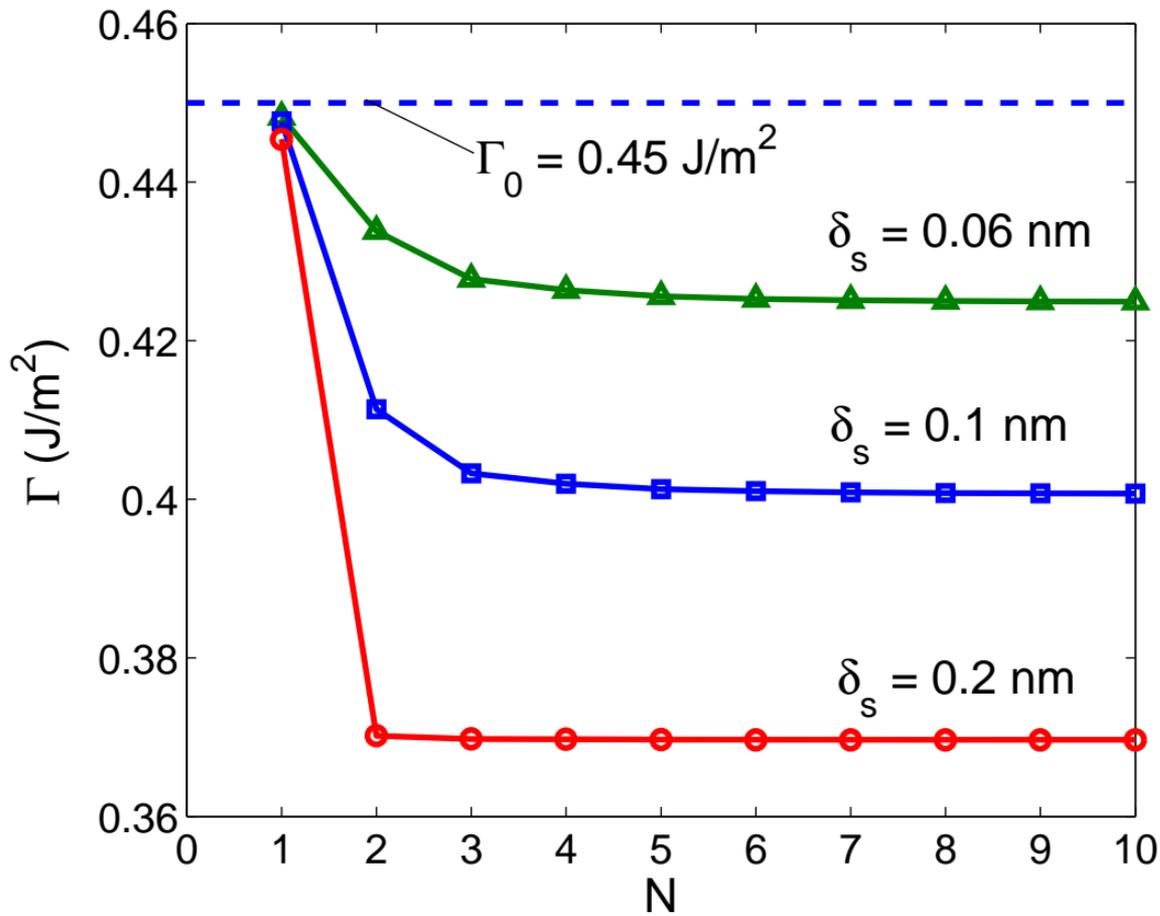